\documentclass[a4paper,11pt,onecolumn,unpublished]{quantumarticle}
\pdfoutput=1

\usepackage[utf8]{inputenc}
\usepackage[colorlinks,pdfusetitle]{hyperref}
\usepackage[numbers]{natbib}

\usepackage{amsmath,amsfonts,amssymb,mathtools}
\usepackage{physics}
\usepackage[group-minimum-digits=0,table-figures-decimal=0,table-number-alignment=center]{siunitx}
\usepackage{quantikz}

\usepackage{amsthm}
\newtheorem{theorem}{Theorem}
\theoremstyle{definition}
\newtheorem{definition}{Definition}

\theoremstyle{remark}

\usepackage{listings}

	\def\equationautorefname~#1\null{Eq.\,(#1)\null}%

\usepackage{dsfont}
\newcommand{\R}{\mathds{R}}

\newcommand{\Z}{\mathds{Z}}

\newcommand{\C}{\mathds{C}}
\newcommand{\D}{\mathds{D}}

\newcommand{\ceil}[1]{{\ensuremath{\left\lceil#1\right\rceil}}}

\newcommand{\identite}{\ensuremath{\mathds{1}}} 

\begin{document}


\title{Provably optimal exact gate synthesis from a discrete gate set}


\author{Élie Gouzien}
\orcid{0000-0002-8209-0681}
\email[]{elie.gouzien@alice-bob.com}
\affiliation{Alice \& Bob, 49 boulevard du Général Martial Valin, \num[detect-all]{75015} Paris, France}
\affiliation{Université Paris--Saclay, CEA, CNRS, Institut de Physique Théorique, \num{91191} Gif-sur-Yvette, France}
\author{Nicolas Sangouard}
\orcid{0000-0002-3136-0266}
\homepage[]{https://quantum.paris}
\affiliation{Université Paris--Saclay, CEA, CNRS, Institut de Physique Théorique, \num{91191} Gif-sur-Yvette, France}

\makeatletter
\hypersetup{pdfauthor={Élie Gouzien and Nicolas Sangouard}}
\makeatother

\date{March 19, 2025}

\begin{abstract}
We propose a method for exact circuit synthesis using a discrete gate set, as required for fault-tolerant quantum computing.
Our approach translates the problem of synthesizing a gate specified by its unitary matrix into a boolean satisfiability (SAT) instance.
It leverages the algebraic properties of the coefficients of the matrices that constitute the gate set, enabling the transformation of the constraint of equality between complex matrices into a boolean expression to satisfy.
For a specified number of gates, the SAT solver finds a circuit implementing the target or proves that none exist.
Optimality of the number of gates is proven by iterating on the number of gates.
We also propose some extensions of the method, for example, handling ancillary qubits, post-selection, or classical feedback.
The time-to-solution scales double-exponentially with the number of qubits, making it impractical for large circuits.
However, since many quantum algorithms rely on small, frequently used subcircuits, and because of the intrinsic value of having a provably optimal circuit synthesis, we believe that our tool are valuable for quantum circuit design.

\end{abstract}

\maketitle


\section{Introduction}
Fault-tolerant quantum computers are extremely arduous to build, and the day they are available at a size relevant to run a useful computation, there will be no room for wasting logical qubits or gates.
Hence, it is important to ensure that their resources are best used.
Fault-tolerant quantum algorithm compilation aims at addressing this challenge, and several approaches are being investigated.
The most basic one consists on using handcrafted expansion rules to replace a high-level function by lower-level ones.
This can be supplemented by circuit-rewriting rules, eventually obtained by exhaustive search, that aim to apply simplifications.
Here we investigate an alternative solution that is not often considered, because it is intrinsically not scalable to a large number of qubits: the synthesis of a circuit from a gate specified by its unitary operator.

In practice, algorithms are built on top of a limited set of small primitives repetitively used, and any improvement on a small subroutine can result in an important gain at the application level.
For example, the cost of Shor’s algorithm is dominated by modular exponentiation, which is typically decomposed into a sequence of multiplications.
These, in turn, are synthesized as a series of additions, each further broken down into a sequence of 3-bit adders~\cite{Ekeraa2019Howfactor2048,SangouardPRL2021Factoring2048bit,SangouardPRL2023PerformanceAnalysisRepetition,Moulton2004newquantumripple}.
Improving the 3-bit adder improves in the same proportion the resources required to run the whole Shor algorithm.

Synthesis algorithms that have been designed to prove universality are able to fulfill the task of generating a circuit implementing a given unitary matrix~\cite{WeinfurterPRA1995Elementarygatesquantum,NielsenQIC2006SolovayKitaevalgorithm}, but optimality of the synthesized circuit has never been their goal.
More recent works have focused on the algebraic properties of the coefficient of matrices to propose efficient decomposition algorithms~\cite{MoscaQIaC2013Fastefficientexact,Selinger2014Optimalancillafree,SelingerPRA2013Exactsynthesismultiqubit,DrechslerQIP2020Advancedexactsynthesis}.
Meet-in-the-middle algorithms have also been proposed~\cite{RoettelerIToCDoICaS2013MeetMiddleAlgorithm,MoscaQSaT2016Parallelizingquantumcircuit}.
In this article, we present a method to map the synthesis problem with a fixed number of gates into a boolean satisfiability problem (SAT).
This allows us to benefit from the performances of SAT solvers.
It also enables leveraging algorithmic proofs to establish the non-existence of a solution.
By iterating on the number of gates, we can hence prove the optimality of a given decomposition.

We start by exposing how to formulate the synthesis problem as a SAT one.
The following section explains how equality between complex matrices can be encoded into binary clauses.
We then discuss the implementation and performance of the method.
A section is also dedicated to discussing variations of the method, that allows to handle slightly different problems.
The code associated with this article is available at \url{https://github.com/ElieGouzien/quatum_gate_sat_synthesis}.

\section{Unitary synthesis as a SAT problem}
\subsection{Short introduction to SAT}
A boolean satisfiability problem (SAT) consists of finding an assignment of boolean variables so that a given logical expression is true, or proving that no such assignment exists.
The expression is usually expected in a conjunctive normal form, i.e., a collection of clauses linked by AND, each clause being the composition with OR of either a variable either its negation.
Solving SAT problems has stimulated a wide community that  has developed algorithms, refined heuristics, and implemented solvers since years.
A competition is annually held: \url{https://satcompetition.github.io/}.
Despite being in general NP-complete, with algorithms having an exponential complexity, the base of this exponential is rather low, being $1.307$ for modern algorithms~\cite{Zwick2019FasterkSAT}.
In addition, SAT solvers arriving at the top of the competition have been particularly well implemented and show themselves practical, including on quite large problems. 

Mapping a problem to a SAT one is an easy way to benefit from the advanced branch-and-bound algorithms of the SAT solvers.
Beyond all the applications of SAT solvers, note that in quantum information processing SAT solvers have also been proposed for Clifford circuit synthesis~\cite{Burgholzer2023DepthOptimalSynthesis,Wille2022TowardsSATEncoding}, circuit rewriting~\cite{DeMicheli2018SATbasedCNOT}, reversible circuit synthesis~\cite{Drechsler2007Exactsatbased,DrechslerIToCDoICaS2009ExactMultipleControl,Drechsler2012ExactSynthesisToffoli}, lattice surgery schedule~\cite{Gidney2024SATScalpelLattice} or code design~\cite{VuillotNC2025LDPCcatcodes}.

\subsection{Problem definition}
We consider the following problem.
We fix a number of qubit $n$.
A target unitary matrix $U$ of shape $2^n \times 2^n$ is given.
We have an indexed gate set ${\{G_j\}}_j$, where each gate $G_j$ is a $2^n \times 2^n$ unitary matrix.
Such gate set is usually obtained by considering some few-qubits gates, e.g.\@ Paulis, $H$, CNOT, $T$, applied to different qubits.
Note that CNOT gate with qubit 1 as control and qubit 3 as target is considered a different gate than CNOT with qubit 0 as control and qubit 1 as target.
Connectivity constraints can be enforced by limiting ${\{G_j\}}_j$ to feasible gates.

The problem consist in finding a circuit implementing exactly $U$ in $d$ gates ($d$ for depth), or proving that no such circuit exits.
The circuit is defined by specifying which gate from ${\{G_j\}}_j$ is applied at each step.

\subsection{Unitary synthesis as a SAT problem}
The main idea is to define a set of binary decision variables, $x_{i, j}$ where $i$ is the index of the step considered (or depth) and $j$ the identifier of the gate.
The gate is applied when $x_{i, j} = 1$ and not when $x_{i, j} = 0$.
With that encoding, the conditions for actually implementing the gate $U$ while having selected exactly one gate for each time step are:
\begin{subequations}\label{eq:sat_formulation}
\begin{gather}
\forall i,\ \exists! j \text{ such that } x_{i, j} = 1 \text{ and } \forall j' \neq j,\ x_{i, j'} = 0 \label{eq:sat_formulation:unique_gate} \\
U = \prod_{i=0}^{d-1} \left[\sum_{j} x_{i,j} G_j \right] \label{eq:sat_formulation:equality}
\end{gather}
\end{subequations}
where the order of the product is ``smaller $i$ on the right (inner), larger $i$ on the left (outer)''.

The exact unicity condition \autoref{eq:sat_formulation:unique_gate} can be expressed in normal form for a given $i$ as:
\begin{equation}
\left[\bigvee_{j} x_{i, j}\right] \land \left[\bigwedge_{j_1, j_2} \neg x_{i, j_1} \lor \neg x_{i, j_2}\right]
\end{equation}

Note that \autoref{eq:sat_formulation:equality} is an equality between complex matrices, and can't be natively encoded into a boolean problem.
This issue is addressed in the next section.

\subsection{Minimal circuit}
The problem formulated above is to decide if there exists a circuit of $d$ gates implementing $U$, and showing such a circuit when it exists.
By solving iteratively this problem for ascending $d$ starting at $1$, the first solution found is guaranteed to be the smallest circuit.

\section{From complex to binary variables}
\subsection{Integers ring extensions}
Let us introduce some extensions of integer numbers.
\begin{definition}
$\Z[x]$ is the ring generated by the integers and $x$ (the ring operations being derived from the ones of $\C$).
In particular:
\begin{itemize}
\item $\Z[i] = \{a + b i \mid a, b \in \Z\}$, ring of Gaussian integers;
\item $\Z[\sqrt{2}] = \{a + b \sqrt{2} \mid a, b \in \Z\}$, ring of integers extended with $\sqrt{2}$ (or quadratic integers with radicand 2);
\item $\Z[\sqrt{2}, i] = \{a + b i + c \sqrt{2} + d i \sqrt{2} \mid a, b, c, d \in \Z\}$;
\item $\D = \Z[\frac{1}{2}] = \{\frac{a}{2^k} \mid a \in \Z, k \in N\}$, ring of dyadic fractions;
\item $\D[\sqrt{2}] = \Z[\frac{1}{\sqrt{2}}] = \{a + b \sqrt{2} \mid a, b \in \D\}$;
\item $\D[e^{i\frac{\pi}{8}}] = \Z[\frac{1}{\sqrt{2}}, i] = \{a + b i + c \sqrt{2} + d i \sqrt{2} \mid a, b, c, d \in \D\}$;
\end{itemize}
\end{definition}
Those standard definitions are also reminded in~\cite[Def.\,3.1]{Selinger2014Optimalancillafree}, with additional details and other rings (e.g.\@ cyclotomic integers).

To be more explicit on $\Z[\sqrt{2}, i]$ operations, let $u = a + b i + c \sqrt{2} + d i \sqrt{2}$ and $u' = a' + b' i + c' \sqrt{2} + d' i \sqrt{2}$.
We have:
\begin{subequations}
\begin{align}
u + u' &= (a+a') + (b+b')i + (c+c')\sqrt{2} + (d+d')i\sqrt{2} \\
u u' &= \begin{multlined}[t]
		\left[aa' - bb' + 2(cc' - dd')\right]
		+ \left[ab' + ba' + 2(cd'+dc')\right] i \\
		+ \left[ac' + ca' - bd' - db'\right]\sqrt{2}
		+ \left[cb' + ad' + bc' + da'\right] i \sqrt{2}
	\end{multlined}
\end{align}
\end{subequations}

\subsection{Algebraic properties of Clifford+$T$ matrices}
As remarked in previous works~\cite{MoscaQIaC2013Fastefficientexact,SelingerPRA2013Exactsynthesismultiqubit,Selinger2014Optimalancillafree,Wetering2024PicturingQuantumSoftware}, all the unitaries from the gate set
\begin{equation*}
\text{CNOT} = \mqty(1 & 0 & 0 & 0 \\ 0 & 1 & 0 & 0 \\ 0 & 0 & 0 & 1 \\ 0 & 0 & 1 & 0)
\qquad
H = \frac{1}{\sqrt{2}} \mqty(1 & 1 \\ 1 & -1)
\qquad
T = \mqty(1 & 0 \\ 0 & e^{i \frac{\pi}{4}})
\qquad
e^{i \frac{\pi}{4}} \identite
\end{equation*}
have their coefficients in the ring $\Z[\frac{1}{\sqrt{2}}, i]$.
They even obtained results that all gates with coefficients in $\Z[\frac{1}{\sqrt{2}}, i]$ can be decomposed exactly on this gate set, with explicit algorithms doing that~\cite{MoscaQIaC2013Fastefficientexact,SelingerPRA2013Exactsynthesismultiqubit}.

\subsection{From dyadic to integers} 
Let us remind that we aim to obtain a SAT formulation of the synthesis problem, and that the total number of gates $d$ is a given number.
We want to work with variables that can take only finite (also possibly exponentially large) numbers of values.
The first step is to limit ourselves to extended integers $\Z[\sqrt{2}, i]$.

For that, notice that up to a prefactor $\frac{1}{2}$, the matrices of our gate set have their coefficients in $\Z[\sqrt{2}, i]$:
\begin{gather*}
\text{CNOT} = \frac{1}{2}\mqty(2 & 0 & 0 & 0 \\ 0 & 2 & 0 & 0 \\ 0 & 0 & 0 & 2 \\ 0 & 0 & 2 & 0)
\qquad
H = \frac{1}{2} \mqty(\sqrt{2} & \sqrt{2} \\ \sqrt{2} & -\sqrt{2})
\qquad
T = \frac{1}{2} \mqty(2 & 0 \\ 0 & \sqrt{2}+i \sqrt{2}) \\
e^{i \frac{\pi}{4}} = \frac{\sqrt{2} + i\sqrt{2}}{2}
\end{gather*}
Said otherwise, with ${\{G_j\}}_j$ still denoting our gate set, the coefficients of all the matrices $2 G_j$ are in $\Z[\sqrt{2}, i]$.
Note that we can check, by enumerating them, that all the 1-qubit Clifford gates also follow this property.

Taking into account this remark, we can rewrite \autoref{eq:sat_formulation:equality} as:
\begin{equation}\label{eq:equality_integer}
2^d U = \prod_{i=0}^{d-1} \left[\sum_{j} x_{i,j} 2 G_j \right]
\end{equation}
This equation proves that $U$ can be exactly synthesized in $d$ gates only if the coefficients of $2^d U$ are in $\Z[\sqrt{2}, i]$.
When it is the case, it reduces the \autoref{eq:sat_formulation:equality} to an equality constraint between integers, and not any more generic complex numbers.

\subsection{From integers to bounded integers}
We now see how to bound the integers.
\begin{definition}\label{def:norm}
Let write a generic number from $\Z[\sqrt{2}, i]$ as $u = a + bi + c\sqrt{2} + d i \sqrt{2}$.
Considering $a$, $b$, $c$ and $d$ as being part of a vector, we can define the following norm:
\begin{equation*}
\norm{u} = a^2 + b^2 + 2c^2 + 2d^2
\end{equation*}
\end{definition}
The fact that it is a norm directly comes from the fact that it is the Euclidean norm up to a rescaling factor on $c$ and $d$  (the matrix defining the corresponding scalar product is diagonal).
Note that the triangular inequality follows: $\norm{u + u'} \leq \norm{u} + \norm{u'}$.

All the gates of the Clifford+$T$ gate set are such that $2G_j$ has at most 2 non-zero coefficients on each line.
Either only one coefficient is non-zero, in which case it is $\pm 2$, $\pm i 2$ or $\sqrt{2} \pm i \sqrt{2}$.
Either exactly two coefficients are non-zero, and those coefficients are $\pm\sqrt{2}$, $\pm i\sqrt{2}$ or $1 \pm i$.
Note that this property is preserved when applying tensor product with identity (as required to match the definition of $G_j$).

Using that $\norm{2 u} = \norm{2 i u} = 2 \norm{u}$, $\norm{\sqrt{2} u} = \norm{i \sqrt{2} u} = \norm{(1 \pm i)u} = \sqrt{2} \norm{u}$ and $\norm{(\sqrt{2} \pm i \sqrt{2}) u}^2 = 2 \norm{u}$ (see \autoref{appendix:norms}), we conclude that if a unitary matrix $U$ with coefficients in $\Z[\sqrt{2}, i]$ has all coefficients bounded by a constant $k$ (for the norm $\norm{\bullet}$), for all $j$ the matrix $2G_j U$ has all coefficients bounded by $2 \sqrt{2} k$.
Using this recursively, we get that after $d$ gates from Clifford+$T$, the right-hand side of \autoref{eq:equality_integer} has its coefficients bounded by ${\left(2\sqrt{2}\right)}^d = 2^{1.5 d}$ (in the sense of the norm $\norm{\bullet}$ on the full coefficient)\footnote{We do not claim this bound to be tight. The unitarity of the target matrix, as well as numerical tests, would suggest a bound of $2^d$ (not proven here).}.

\subsection{From bounded integers to binary variables}

From the \autoref{def:norm}, when a number $u = a + bi + c\sqrt{2} + d i \sqrt{2}$ is such that $\norm{u} \leq k$, we obtain that the integers $a$, $b$, $c$, $d$ verify
$-k \leq x \leq k$, with $x$ any of those integers.
As we have shown previously that $k = 2^{1.5 d}$ when considering $2^d U$, by taking a margin of 1 bit to make the inequality strict and 1 additional bit to be a sign bit, we can conclude:
\begin{theorem}\label{thm:bound}
We consider a gate set such that for each gate, 2 times the unitary matrices have coefficients in $\Z[\sqrt{2}, i]$, with either one non-zero coefficient of type $\pm 2$, $\pm 2 i$ or $\pm \sqrt{2} \pm i \sqrt{2}$ or two non-zero coefficients of type $\pm \sqrt{2}$, $\pm i \sqrt{2}$ or $1 \pm i$.
A circuit generated from $d$ gates taken from such gate set implements a unitary matrix $U$ such that $2^d U$ has coefficients in $\Z[\sqrt{2}, i]$ whose integers can be represented on $\ceil{1.5 d} + 2$ bits.
\end{theorem}
Note that 1-qubit Clifford + CNOT + $T$ verify the property needed in the theorem.

Now that we have shown that the \autoref{eq:sat_formulation:equality} can be rewritten as an equality between bounded integers, the only step left is to realize that it is possible to consider the bits of integers as decision variables of the SAT problem, and to translate the arithmetic expressions as constraints (sometime at the cost of adding ancillary variables).
This is done by bit-blasting the integers, what is possible by realizing that arithmetic operations can be written as logical expressions on the bits composing each integer~\cite{Fredrikson2022LectureNotesBit,Bjoerner2008Z3EfficientSMT} (see \autoref{appendix:bit-blast}).

Directly expressing \autoref{eq:equality_integer} as a single logical clause would not be possible, as expending the right-hand side would already require $g^d$ ($g$ is the gate set size, and $d$ the number of gates) terms\footnote{That was our first implementation attempt; unsurprisingly it didn't work.}.
To remediate this, we introduce the intermediate unitaries $U_i$:
\begin{equation}
2^i U_i = \prod_{i'=0}^{i} \left[\sum_{j} x_{i',j} 2 G_j \right]
\end{equation}
The coefficients of the $2^i U_i$ are considered as decision variables of the SAT problem, with the set of constraints:
\begin{subequations}
\begin{gather}
2 U_0 = \sum_{j} x_{0,j} 2 G_j \\
\forall i<d,\ 2^{i+1} U_{i+1} = \left[\sum_{j} x_{i+1,j} 2 G_j \right] 2^i U_i \label{eq:sat_propagation:general}
\end{gather}
\end{subequations}
With those additional variables, we now have a complete mapping from the synthesis problem to a SAT formulation.

Note that so far we have hidden the fact that \autoref{eq:sat_formulation:equality} requires strict equality between the targeted gate and the unitary matrix of the circuit implementing it.
This can be relaxed by introducing global phase gates in the gate set.
The only phase gates such that $2 G_j$ have coefficients in $\Z[\sqrt{2}, i]$ are the multiples of $e^{i\frac{\pi}{4}}$.
Note that handling an arbitrary phase on the target unitary is not a goal; for example, if $U$ is replaced by $e^{i \frac{\pi}{3}} U$, the coefficients of the new matrix are no more in $\Z[\frac{1}{\sqrt{2}}, i]$, and the approach presented in this article can't be used.

\section{Implementation and performance}
\subsection{Implementation}

For the formulation of the problem, we relied on SymPy~\cite{ScopatzPCS2017SymPysymboliccomputing} and the SMT solver Z3~\cite{MicrosoftResearchZ3,Bjoerner2008Z3EfficientSMT} (for which we did not use the solver itself).
SymPy allows manipulation of human-friendly matrices to define the target and gate set, while Z3 performs the formulation of the SAT problem itself, including the bit-blasting.
The decision variables $x_{i, j}$ are formulated as boolean ones, while all the integer coefficients of the $2^i U_i$ are declared to Z3 as \lstinline|BitVec|, and only latter are blasted to booleans.
In practice, the bound from \autoref{thm:bound} can be used with $i$ instead of $d$ for each $U_i$\footnote{Experiments seems to suggest that the bound of \autoref{thm:bound} is not tight and that $d+2$ could be enough. We let this question open (anyway, SAT solvers are efficient at removing pointless variables).}.
To avoid complications in the definition of ``number of gates'', we did not add global phase in the gate set, but treated them independently.
The SAT problem is then exported and saved in conjunctive normal form in a file under the DIMACS format.

The SAT solver Kissat~\cite{BiereKissatSATSolver,Pollitt2024CaDiCaLGimsatulIsaSAT}, that won the recent SAT competitions, is then used to solve the SAT problem.
Helper function then eases reading the result file output by Kissat.

Our code is published at \url{https://github.com/ElieGouzien/quatum_gate_sat_synthesis}.

\subsection{Performances}
In the following discussion, we consider only the time spent in the SAT solver, as the problem formulation is usually faster, of polynomial complexity and has a lot of room for improvement.
Kissat was run on a standard desktop computer; it is a single-thread program.
The code presented above is typically able to find a solution, or prove there are none, in a few days for problems with 3 qubits and 10--15 gates, but this is really problem-dependent.
When the problem is particularly simple, it is possible to go well beyond.
For example, the preparation of GHZ states on 7 qubits is found in less than \SI{10}{\minute} (but the formulation is particularly long).

\begin{figure}[h]
\centering
\begin{quantikz}[row sep=1.5em, column sep=2ex]
	 & \qw      & \qw      & \ctrl{2} & \qw              & \qw      & \qw      & \ctrl{2} & \qw              & \qw      & \ctrl{1} & \qw              & \ctrl{1} & \gate{T} & \qw \\
	 & \qw      & \qw      & \qw      & \qw              & \ctrl{1} & \qw      & \qw      & \qw              & \ctrl{1} & \targ{}  & \gate{T^\dagger} & \targ{}  & \gate{T} & \qw \\
	 & \gate{H} & \gate{T} & \targ{}  & \gate{T^\dagger} & \targ{}  & \gate{T} & \targ{}  & \gate{T^\dagger} & \targ{}  & \gate{H} & \qw              & \qw      & \qw      & \qw
\end{quantikz}
\caption{Toffoli decomposition obtained in about 6 days by our method, with $g=15$, $d=15$.}\label{fig:toffoli}
\end{figure}
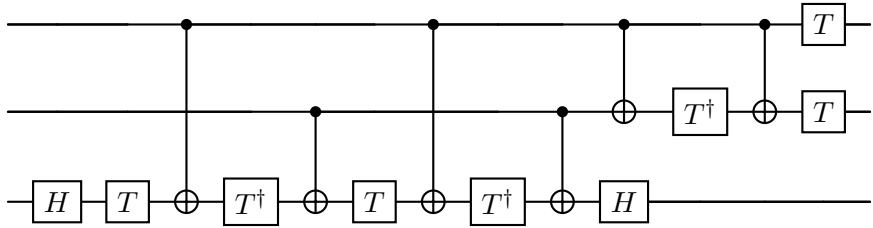

A first example of a typical problem to address with such a tool is the decomposition of the Toffoli gate on a Clifford+$T$ gate set.
With the method of this paper, we obtained the result presented on \autoref{fig:toffoli} (see also \href{https://algassert.com/quirk#circuit=%7B%22cols%22%3A%5B%5B1%2C1%2C%22H%22%5D%2C%5B1%2C1%2C%22Z%5E%C2%BC%22%5D%2C%5B%22%E2%80%A2%22%2C1%2C%22X%22%5D%2C%5B1%2C1%2C%22Z%5E-%C2%BC%22%5D%2C%5B1%2C%22%E2%80%A2%22%2C%22X%22%5D%2C%5B1%2C1%2C%22Z%5E%C2%BC%22%5D%2C%5B%22%E2%80%A2%22%2C1%2C%22X%22%5D%2C%5B1%2C1%2C%22Z%5E-%C2%BC%22%5D%2C%5B1%2C%22%E2%80%A2%22%2C%22X%22%5D%2C%5B1%2C1%2C%22H%22%5D%2C%5B%22%E2%80%A2%22%2C%22X%22%5D%2C%5B1%2C%22Z%5E-%C2%BC%22%5D%2C%5B%22%E2%80%A2%22%2C%22X%22%5D%2C%5B%22Z%5E%C2%BC%22%2C%22Z%5E%C2%BC%22%5D%5D%2C%22init%22%3A%5B%22%2B%22%2C%22%2B%22%5D%7D}{it on quirk}).
It has a lot in common but is slightly different from the standard decomposition~\cite[Fig.\,4.9]{Chuang2010QuantumComputationQuantum}.
Here we considered the gate set $\{H_j, T_j, T^\dagger_j,\ 0 \leq j < 3  \} \cup \{\text{CNOT}_{j, j'},\ 0 \leq j, j' < 3,\ j \neq j'\}$, that is of cardinal 15.
It took around 6 days to be obtained.
As a reference, the search space of the original problem has a size of ${15}^{15} \approx 4.4 \times {10}^{17}$.
To give a hint on how efficient Kissat is at avoiding a complete exploration of the search space, note that the $x_{i, j}$ can take $2^{15 \times 15} \approx 5.4 \times {10}^{67}$ different values.
The input of Kissat is a SAT problem with \num{1104153} variables and \num{6607851} clauses.

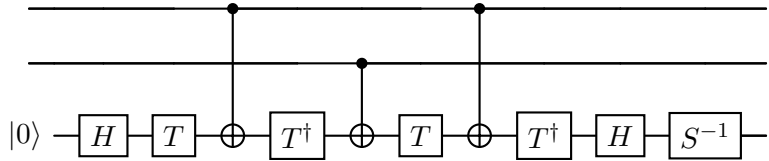
\begin{figure}[h]
\centering
\begin{quantikz}[row sep=1.5em, column sep=2ex]
	 & \qw                                & \qw      & \qw      & \ctrl{2} & \qw              & \qw      & \qw      & \ctrl{2} & \qw              & \qw      & \qw           & \qw \\
	 & \qw                                & \qw      & \qw      & \qw      & \qw              & \ctrl{1} & \qw      & \qw      & \qw              & \qw      & \qw           & \qw \\
	 & \lstick{$\ket{0}$}\wireoverride{n} & \gate{H} & \gate{T} & \targ{}  & \gate{T^\dagger} & \targ{}  & \gate{T} & \targ{}  & \gate{T^\dagger} & \gate{H} & \gate{S^{-1}} & \qw
\end{quantikz}
\caption{%
	AND gate decomposition obtained from our solver.
	It is identical from the original proposition of AND gate~\cite[Fig.\,4]{NevenPRX2018EncodingElectronicSpectra}, up to the permutation of the two controls.
}\label{fig:AND}
\end{figure}

Another example is the AND gate (Toffoli targeting a qubit prepared in $\ket{0}$).
The translation to SAT problem requires one of the extensions presented in the next section.
The result is shown in \autoref{fig:AND} (also accessible \href{https://algassert.com/quirk#circuit=%7B%22cols%22%3A%5B%5B1%2C1%2C%22H%22%5D%2C%5B1%2C1%2C%22Z%5E%C2%BC%22%5D%2C%5B%22%E2%80%A2%22%2C1%2C%22X%22%5D%2C%5B1%2C1%2C%22Z%5E-%C2%BC%22%5D%2C%5B1%2C%22%E2%80%A2%22%2C%22X%22%5D%2C%5B1%2C1%2C%22Z%5E%C2%BC%22%5D%2C%5B%22%E2%80%A2%22%2C1%2C%22X%22%5D%2C%5B1%2C1%2C%22Z%5E-%C2%BC%22%5D%2C%5B1%2C1%2C%22H%22%5D%2C%5B1%2C1%2C%22Z%5E-%C2%BD%22%5D%5D%2C%22init%22%3A%5B%22%2B%22%2C%22%2B%22%5D%7D}{on quirk}).
In this example, the gate set was larger, with 33 possible gates: $\{X_j, Y_j, Z_j, H_j, S_j, S^\dagger_j, T_j, T^\dagger_j,\ 0 \leq j < 3  \} \cup \{\text{CNOT}_{j, j'},\ 0 \leq j, j' < 3,\ j \neq j'\} \cup \{{\text{C}Z}_{j, j'},\ 0 \leq j < j' < 3\}$.
The number of gates in the smallest solution is 10, and the computation took around 4 days.
For reference, $33^{10} \approx 1.5 \times {10}^{15}$ and $2^{33 \times 10} \approx 2.2 \times {10}^{99}$.

\section{Variations}

We have explored different variations of the problem formulation to allow handling slightly different problems.
In this section, we briefly describe a certain number of them.

\subsection{Bounds on gate number by type}
It can be useful to set a limit on the number of one type of gates, for example, non-Clifford gates.
This is done by imposing a cardinality constraint on a selection of $x_{i, j}$.
It is formulated in Z3 by using and \lstinline|AtMost| constraint.
Z3 handles the translation of such a pseudo-boolean constraint into a boolean one, similarly to what is done to impose that only one gate is picked for each time-step.

Similarly to the main problem formulation, iterating on such a bound allows proving the optimality of a circuit in terms of the number of one type of gate.

\subsection{Dirty ancilla}
A dirty ancilla is a qubit not part of the input qubits of the target unitary $U$, that is in an unspecified state.
It can be used, but needs to be restored to its original state (and not entangled with the outputs) at the end of the circuit.
Finding a circuit that implements a unitary $U$ with the help of a dirty ancilla is equivalent to searching for a circuit that implements $U \otimes \identite$.

\subsection{Fixed input and clean ancilla}\label{extensions:fixed}
Some gates, such as the AND gate ($\forall a, b \in \{0, 1\},\ \ket{a}\ket{b} \mapsto \ket{a}\ket{b}\ket{ab}$) have more output than input.
They are described by a rectangular matrix, with fewer columns than lines.
Said differently, we can consider that some inputs have a mandatory fixed value.
It is handled by removing the undesired columns in all the intermediate matrix $U_i$.

This is also the case when dealing with clean ancilla initialized in $\ket{0}$: we treat this by searching a circuit for $U \otimes \identite$, with the columns corresponding to the value $\ket{1}$ for the ancillary qubit removed.

\subsection{Post-selection and classical feedback}
Handling post-selection after a measurement in the computational basis is done by removing from the unitary the lines corresponding to the selected-out states.
Renormalization is also needed.
The approach we have followed is to treat only the case where the measurement results have both probability $\frac{1}{2}$, hence the renormalization factor does not depend on the circuit.
Despite being restrictive, it corresponds to the case of measurement-based uncomputing, or of teleportation (0 and 1 can also be handled, but we don't need post-selection on a deterministic result).
More refined approaches are certainly possible but are left for future work.

Classical feedback refers to the application of gates conditioned on the result of a measurement.
In our case, the step at which the measurement happens is decided in advance, similarly to the total number of gates.
In practice, we do not change the shape of the intermediate matrices associated with the circuit after that step, but the classical controls on the qubits that should have been measured are replaced by quantum ones, while forbidding operations that don't map the computational basis to itself.
This is formally equivalent to pushing the measurements at the end of the circuit, while maintaining the constraint that only classically controlled gates can be used once a qubit is measured.
We force all the possible measurement outcomes (formally treated as  different post-selections) to implement the same target gate on the other qubits.
In practice, it is implemented by defining the gate set as containing all the possible gates, and adding constraints to forbid their use where they are not allowed.

\begin{figure}[h]
	\centering
	\begin{quantikz}[row sep=1.5em, column sep=2ex]
		& \qw                                & \qw      & \qw      &          & \qw              & \ctrl{3} & \qw      &          & \qw              & \qw      &           & \qw           & \qw      &                    &                  \\
		& \qw                                & \qw      & \qw      & \ctrl{2} & \qw              &          & \qw      & \ctrl{2} & \qw              & \qw      &           & \qw           & \qw      &                    &                  \\
		&                                    &          &          &          &                  &          &          &          &                  &          & \targ{}   &               &          &                    &                  \\
		& \lstick{$\ket{0}$}\wireoverride{n} & \gate{H} & \gate{T} & \targ{}  & \gate{T^\dagger} & \targ{}  & \gate{T} & \targ{}  & \gate{T^\dagger} & \gate{H} & \ctrl{-1} & \gate{S^{-1}} & \gate{H} & \rstick{$\bra{0}$} & \wireoverride{n}
	\end{quantikz}
	\caption{%
		Toffoli gate, with an ancilla and post-selection on $\ket{0}$, as found by our solver.
		It corresponds to the implementation of the Toffoli gate through a AND gate, a CNOT and measurement-based uncomputing, for which an additionnal classically-controlled C$Z$ on the two first qubits is required to handle the measurement outcome 1.
	}\label{fig:toffoli_ancilla}
\end{figure}
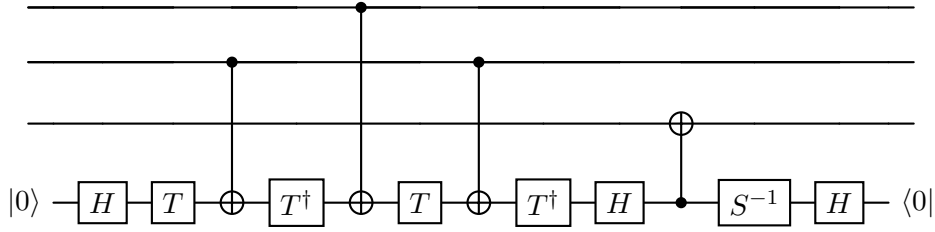

An example is shown in \autoref{fig:toffoli_ancilla} (also visible \href{https://www.algassert.com/quirk#circuit=%7B%22cols%22%3A%5B%5B1%2C1%2C1%2C%22H%22%5D%2C%5B1%2C1%2C1%2C%22Z%5E%C2%BC%22%5D%2C%5B1%2C%22%E2%80%A2%22%2C1%2C%22X%22%5D%2C%5B1%2C1%2C1%2C%22Z%5E-%C2%BC%22%5D%2C%5B%22%E2%80%A2%22%2C1%2C1%2C%22X%22%5D%2C%5B1%2C1%2C1%2C%22Z%5E%C2%BC%22%5D%2C%5B1%2C%22%E2%80%A2%22%2C1%2C%22X%22%5D%2C%5B1%2C1%2C1%2C%22Z%5E-%C2%BC%22%5D%2C%5B1%2C1%2C1%2C%22H%22%5D%2C%5B1%2C1%2C%22X%22%2C%22%E2%80%A2%22%5D%2C%5B1%2C1%2C1%2C%22Z%5E-%C2%BD%22%5D%2C%5B1%2C1%2C1%2C%22H%22%5D%2C%5B1%2C1%2C1%2C%22%7C0%E2%9F%A9%E2%9F%A80%7C%22%5D%2C%5B%22%E2%80%A2%22%2C%22%E2%80%A2%22%2C%22Chance%22%5D%5D%2C%22init%22%3A%5B%22%2B%22%2C%22%2B%22%5D%7D}{on quirk}), where we imposed the post-selection on the ancillary qubit to end in $\ket{0}$.
The solution is found in little more than one hour, with $g=10$ (reduced gate set) and $d=12$.
Implementing a scheme for classical feedback and also enforcing the post-selection on $\ket{1}$ to give a Toffoli gate would have force the solver to add a classically controlled C$Z$ on the two first qubits.

\subsection{Specification by state mapping}
It can happen that instead of a unitary matrix, only the mapping of a few states is given, for example in the context of the implementation of logical gates on an error-correcting code.
This is equivalent to the case of \autoref{extensions:fixed}, except that the specified states can be outside the computational basis.
This could be handled by starting from the matrix encoding those states instead of $\identite$.
We decided to simply concatenate the constraints that each state is mapped to the correct one (it is equivalent).

\subsection{Richer gate set}
We presented the method in the case where for all $j$, $2 G_j$ has coefficients in $\Z[\sqrt{2}, i]$.
The method remains applicable to more general cases, as long as one can find a relevant prefactor and finite extension of the integers.

For example, to handle $\sqrt{T}$, we can work on $\Z\left[\sqrt{2}, \sqrt{2 + \sqrt{2}}\right]$ (more precisely its complex version).
This is done by parametrizing the ring as $a + b\sqrt{2} + c\sqrt{2+\sqrt{2}} + d\sqrt{2}\sqrt{2+\sqrt{2}}$, with $a, b, c, d \in \Z$.

Note that it is not straight forward  to identify from a generic expression if it is part of the ring, and to get the coefficients.
For example, $\sqrt{2-\sqrt{2}} =  \sqrt{2}\sqrt{2+ \sqrt{2}} - \sqrt{2 + \sqrt{2}}$.
This is typically handled by working on minimal polynomials.
In practice, we rely on SymPy's handling of algebraic fields.

\subsection{Alternative formulation with mixed-integer linear programming}

Note that \autoref{eq:sat_formulation:equality} is expressed as an equality constraint, but the problem can also be reformulated as an optimization one.
More precisely, the objective is to minimize a distance between the right-hand side of \autoref{eq:sat_formulation:equality} and $U$.
Similarly as for the SAT formulation, we introduce a sequence of matrices $U_i$ representing the unitary implemented after the step $i$ of the circuit.
The succession constraint (replacing \autoref{eq:sat_propagation:general}) can be encoded using:
\begin{equation}
\forall i, j,\ x_{i+1, j} \implies U_{i+i} = G_j U_{i}
\end{equation}
To see each such constraint (for given $i$ and $j$) as a linear one, we note that an equality constraint can always be decomposed as two inequality constraints.
Then, remark that $x \implies a^\dagger u \leq b$, with $x$ a binary variable, $u$ a vector of continuous ones, $a$ a vector of constants, and $b$ a constant, can be rewritten as $a^\dagger u \leq b + C (1-x)$, where $C$ is an arbitrary constant number chosen large enough.

The condition to choose exactly one gate at each step (\autoref{eq:sat_formulation:unique_gate}) can be reformulated as:
\begin{equation}
\forall i,\ \sum_{j} x_{i, j} = 1
\end{equation}

Finally, note that using the Euclidean distance would result in a quadratic objective.
Instead, we use the $\ell_1$ distance, noting that we can enforce $y = \abs{x}$ by adding a binary variable $c$ and imposing $y \geq 0$, $c \implies y = x$, $(1-c) \implies y = -x$.

With those transformations, the problem appears as a mixed-integer linear program.
Efficient solvers exist for such problems, and this formulation has the advantage of providing a way to find an approximate implementation of a gate, and not only an exact one.
In addition, optimality of the approximation (for the specified number of steps $d$) can be proven by the solver.

Experiments have shown this method to be efficient for gates on very few qubits, especially when there is only one, even for a large number of steps $d$.
However, the SAT formulation and implementation appears largely faster when exact synthesis can be achieved.

\subsection{Classical reversible circuits}
For the synthesis of classical reversible circuits, we used a similar approach, but specialized.
The gate set is NOT ($X$), CNOT and Toffoli, on all possible qubits, and it only maps states from the computational basis to other states of the computational basis.
Handling of the decision variable doesn't change compared with the generic case.
Enforcing to perform the correct function is handled differently: there is no need to represent unitary matrices, and only a vector of boolean variables is defined at each step.
The selected gates impose relations between one step and the following.
We iterate over all possible inputs and impose the correct output for all of them.

Note that the same approach have already been formulated~\cite{Drechsler2007Exactsatbased,Drechsler2012ExactSynthesisToffoli}\footnote{See \url{https://github.com/msoeken/reversible-sota} for an (outdated) state of the art on reversible circuit synthesis, and \url{https://github.com/msoeken/revkit} for an implementation of~\cite{Drechsler2012ExactSynthesisToffoli}.}.

An alternative formulation using quantifiers and relying on the SMT solver of Z3 have also be attempted, but it turned out to be slower despite a very compact formulation.
This idea has also been explored in the past~\cite{GroBe2008QuantifiedSynthesisReversible}.

Typical usage of such code allowed us to discover or validate optimality of the different variants of adders introduced in~\cite{SangouardPRL2023PerformanceAnalysisRepetition}.
It also allows us to conjecture the optimality of the method presented in~\cite{Gidney2024Riseconditionallyclean} for implementing a multi-controlled NOT gate with only one clean ancilla.

\section{Conclusion}

We showed how to map the exact synthesis problem into a SAT one, in the case where the target is specified by a unitary matrix, the number of gates is defined, and the gate set allows to use ring extensions of $\Z$.
By iterating on the number of gates, it allows finding the smallest circuit solution, and proving the optimality.
We also presented a collection of variations, the most important ones being the classical reversible circuit, and the linear programming ones.

Only small circuits can be tackled with this method, but it allowed us to obtain most of the known small circuits (including some that took 20 years for the whole community to be found).
The proof of optimality is also very important, as it helps to conjecture and prevents wasted effort on improving an already optimal circuit.
We now let the reader use it to discover unknown circuits.

\begin{acknowledgments}
We acknowledge Xavier Valcarce and Julian Zivy for stimulating discussions, and Christophe Vuillot for proofreading an earlier version of this manuscript.
This work was partially funded by the European High-Performance Computing Joint Undertaking (JU) under grant agreement No 101018180 and project name HPCQS and by the French National program ``Programme d’investissement d’avenir'', IRT Nanoelec, ANR-10-AIRT-05.
\end{acknowledgments}

\appendix
%
\section{Additional note on norms}\label{appendix:norms}

\subsection{Some explicit norm computations}
Let $u = a + bi + c\sqrt{2} + d i \sqrt{2}$.
Hence, we have:
\begin{gather*}
1 \times u = u \\
i \times u = -b + ai - d\sqrt{2} + c i\sqrt{2} \\
2 \times u = 2a + 2bi + 2c\sqrt{2} + 2d i \sqrt{2} \\
2i \times u = -2b + 2ai - 2d\sqrt{2} + 2c i\sqrt{2} \\
\sqrt{2} \times u = 2c + 2d i + a \sqrt{2} + b i \sqrt{2} \\
i \sqrt{2} \times u = -2d + 2ci - b \sqrt{2} + a i \sqrt{2} \\
(1 \pm i) \times u = (a \mp b) + (\pm a + b)i + (c \mp d)\sqrt{2} + (\pm c+d) i\sqrt{2} \\
(\sqrt{2} \pm i \sqrt{2}) \times u = 2(c \mp d) + 2(\pm c+d) i + (a \mp b) \sqrt{2} + (\pm a+b) i \sqrt{2}
\end{gather*}

From those, and with the norm $\norm{\bullet}$ as defined in \autoref{def:norm}, we have:
\begin{gather*}
\norm{2 u} = \norm{2 i u} = 2 \norm{u} \\
\norm{\sqrt{2} u} = \norm{i \sqrt{2} u} = \sqrt{2} \norm{u} \\
\norm{(1 \pm i)u}^2 = 2(a^2 + b^2) + 4(c^2 + d^2) = 2 \norm{u}^2 \implies \norm{(1 \pm i)u} = \sqrt{2}\norm{u} \\
\norm{(\sqrt{2} \pm i \sqrt{2}) u}^2 = 8(c^2+d^2) + 4(a^2 + b^2) = 4 \norm{u}^2
	\implies \norm{(\sqrt{2} \pm i \sqrt{2}) u} = 2 \norm{u}
\end{gather*}

\subsection{Other norms}
Still using $u = a + bi + c\sqrt{2} + d i \sqrt{2}$, we could have defined the following norms:
\begin{itemize}
\item $\max(\abs{a}, \abs{b}, \abs{c}, \abs{d})$
\item $\max(\abs{a}, \abs{b}, \sqrt(2)\abs{c}, \sqrt(2)\abs{d})$
\item $a^2 + b^2 + c^2 + d^2$
\item ${(a+c\sqrt{2})}^2 + {(b+d\sqrt{2})}^2 = a^2 + 2 c^2 + 2\sqrt{2} ac + b^2 + 2d^2 + 2\sqrt{2}bd$ (standard complex modulus; only a seminorm on $\R^4$)
\end{itemize}
They either don't allow to bound the individual coefficient, or would have given a looser bound.

\section{Bit-blasting}\label{appendix:bit-blast}

An adder between fix-sized integer can be written as a sequence of 3-bit additions.
A 3-bit adder can be written in conjunctive normal form:
writing $a$, $b$, and $c$ its inputs, $r$ the result and $c'$ the output carry,
\begin{align*}
	c' &= a b \oplus b c \oplus a c
		=  (a \land b) \lor (b \land c) \lor (a \land c)
		=  (a \lor b) \land (b \lor c) \land (a \lor c) \\
	r &= a \oplus b \oplus c
		= (a \lor b \lor c) \land (a \lor \neg b \lor \neg c) \land (\neg a \lor b \lor \neg c) \land (\neg a \lor \neg b \lor c)
\end{align*}
Hence, the equality between the sum of two inputs and the result can be enforced as a SAT clause (carries are typically added as ancillary variables).

Multiplications can be decomposed into sequences of additions, and can hence be bit-blasted.

For a more extensive introduction to bit-blasting, see~\cite{Fredrikson2022LectureNotesBit}.

\bibliographystyle{quantum}
\bibliography{article_sat_synthese}

\end{document}